# Effect of nano-C doping on the *in-situ* processed MgB$_2$ tapes

Xianping Zhang[1], Yanwei Ma[1*], Aixia Xu[1], Yulei Jiao[2], Ling Xiao[2], S. Awaji[3], K. Watanabe[3], Huan Yang[4], Haihu Wen[4]

[1] Applied Superconductivity Lab., Institute of Electrical Engineering, Chinese Academy of Sciences, Beijing 100080, China
[2] General Research Institute for Nonferrous Metals, Beijing 100088, China
[3] Institute for Materials Research, Tohoku University, Sendai 980-8577, Japan
[4] Institute of Physics, Chinese Academy of Sciences, Beijing 100080, China

\* E-mail:ywma@mail.iee.ac.cn

**Abstract**. The effect of nano-C doping on the microstructure and superconducting properties of Fe-sheathed MgB$_2$ tapes prepared through the *in-situ* powder-in-tube method was studied. Heat treatment was performed at a low temperature of 650°C for 1 h. Scanning electron microscopy investigation revealed that the smaller grain size of MgB$_2$ in the samples with the C-doping. Further, the a-axis lattice parameter and transition temperature decreased monotonically with increasing doping level, which is due to the C substitution for B. High critical current density J$_C$ values in magnetic fields were achieved in the doped samples because of the very fine-grained microstructure of the superconducting phase obtained with C doping.

## 1. Introduction

The discovery of a superconducting transition at 39 K in MgB$_2$ initiated enormous scientific interest not only in basic physics but also in practical applications [1]. The much higher transition temperatures(T$_C$) of MgB$_2$ caused researchers to assume that MgB$_2$ could be used at elevated temperatures around 20 K as the conductor of a cryogen-free magnet, and high critical current density J$_C$ values of 10$^6$ A/cm$^2$ have been achieved in MgB$_2$ pellets and tapes [2]. However, J$_C$ of MgB$_2$ drops rapidly with increasing magnetic field due to the low upper critical field (H$_{C2}$) and poor flux pinning. As a simple and practical method, chemical doping seems to be the best route to improve flux pinning, as reported so far [3-9]. Of all chemical substitutions that had been under taken, C substitution has been successful in increasing H$_{C2}$ (0) and irreversiblity field(H$_{irr}$) of MgB$_2$ polycrystalline and single crystalline [7-11]. For practical application of superconductors, such as magnets and cables, tapes and wires must be developed. Nevertheless, there have been few reports on C doping in MgB$_2$ tapes. In this work, nano-C doped MgB$_2$ tapes were prepared by the *in-situ* powder-in-tube (PIT) method with different doping levels (0, 2.5, 5, 10, 15 at%). The doping effects on phases, microstructures, and superconductivity of nano-C doped MgB$_2$ tapes were investigated.

## 2. Experimental

Nano-C doped MgB$_2$ tapes were prepared by the *in-situ* PIT method. The sheath materials chosen for this experiment were commercially available pure Fe. Mg (325 mesh, 99.8%), B (amorphous, 99.995%) and C (20-50 nm, amorphous) powders were used as the starting material. The doping levels of nano-C powder were 0, 2.5, 5, 10, 15 at%, respectively. The well-mixed powder was tightly packed into pure iron tubes of 8 mm outside diameter and 1.5 mm wall thickness. The tubes were subsequently swaged and drawn into round wires of about 1.5 mm in diameter. The wires were then rolled to tapes. The final size of the tapes was 3-4 mm in width and about 0.5 mm in thickness. These tapes were sintered at 650°C for 1 h, and then cooled down to room temperature in the furnace. The argon gas was flowed into the furnace during the heat treatment process in order to avoid the oxidation of the samples.

The phase constituent and microstructure of the samples were investigated by using the power X-ray diffraction (XRD), scanning electron microscope (SEM) and energy dispersive spectrometry (EDS). DC magnetization measurement was performed with a superconducting quantum interference device (SQUID) magnetometer. The $T_C$ was defined as the onset temperature at which a diamagnetic signal was observed. Magnetization hysteresis was measured for the rectangular $MgB_2$ core by using a SQUID magnetometer in fields up to 5 T at 5 K and 20 K. The magnetic critical current density was estimated with a Bean model of $J_C = 20 \ M/[a(3b-a)/3b]$, where $a$ and $b$ are the dimensions of the sample perpendicular to the direction of applied magnetic field with $a < b$.

## 3. Results and discussion

Figure 1 shows the XRD patterns of the heat-treated $MgB_2$ tapes doped with different levels of nano-C powder, the XRD pattern of the starting C powder is also show in the figure. The main diffraction peaks for all samples were identified to be the $MgB_2$ phase in all doping level samples, with only a small amount of MgO present (< 5%), consistent with the oxygen peaks in EDS images. We could not observe any carbon peaks in the XRD patterns of all doping level samples, due to the amorphous C powder we used, similar to other C doping results [12]. It is noted that the (110) peak of the nano-C doped samples shifts systematically to higher angles with increasing doping level. This meaning that the shrinkage of a-axis length was happened due to carbon substitution for boron site, very similar to recent results [12]. The changes in lattice parameters indicated the lattice distortion of $MgB_2$, which usually result in an enhancement of flux pinning [3,13]. No change was observed in peak positions of the (002) reflections, suggesting that C doping does not affect the c-axis.

Figure 2 shows the superconducting transition temperatures ($T_C$) for the undoped and doped tape samples measured by SQUID magnetometer. Clearly, the $T_C$ decreases with increasing nano-C doping level. The $T_C$ onset for the undoped sample is ~36.5 K, for the 2.5% nano-C doped sample, the $T_C$ (onset) decreases to ~35 K. But the $T_C$ difference among x=2.5%, 5%, 10% is very small, within 1 K. The reaction between Mg and C or substitution B by C is proposed to be responsible for the decrease in transition temperature of nano-C doped samples [3].

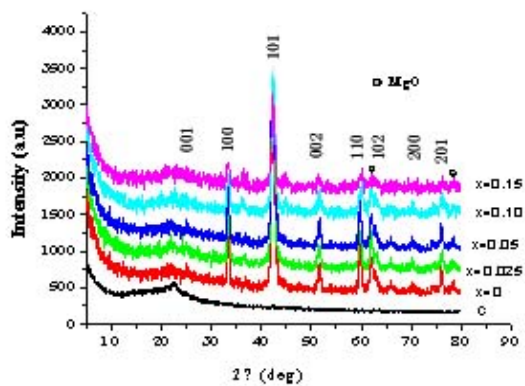

Figure1. XRD patterns of *in-situ* processed nano-C doped $MgB_2$ samples.

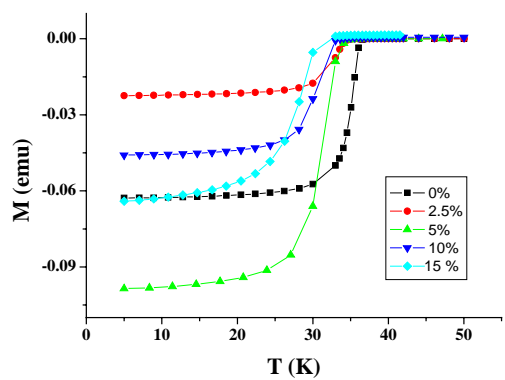

Figure 2. Superconducting transitions of nano-C doped $MgB_2$ tapes.

Figure 3 shows the typical SEM images of fractured core for all doped samples. It is clear that the grain size decreases with the nano-C doping. Thus the fine grain size creates many grain boundaries that may act as effective pinning centers, which will lead to the enhancement of $J_C$-B performance in C-doped samples, as reported previously [14]. In addition, the connection between grains is improved by nano-C doping, hence the enhancement of critical current densities of nano-C doped tape is also expected.

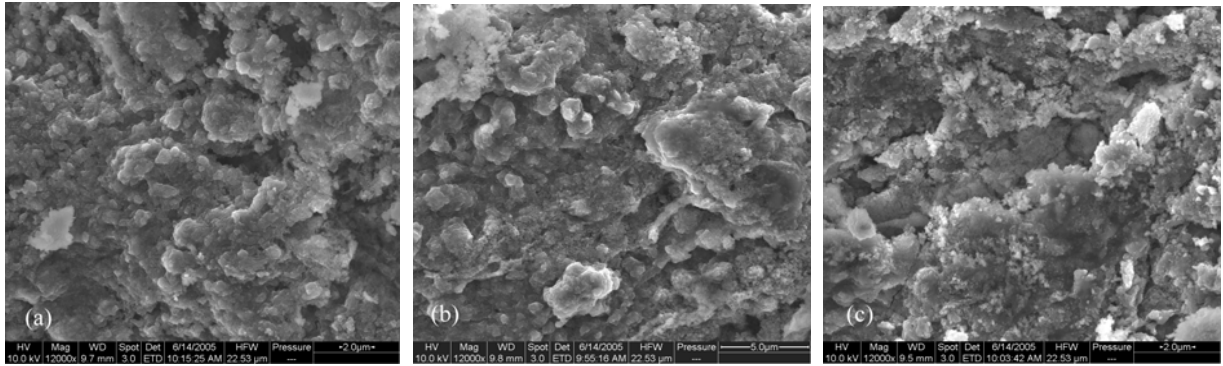

Figure 3. SEM micrographs for nano-C doped $MgB_2$ samples (a) 0%; (b) 5%; (c) 10%.

Figure 4 shows the field dependence of magnetic critical current densities of x=0%, 5%, 10% samples. The magnetic field was applied parallel to the sample surface. The nano-C doped samples revealed a significant improvement in $J_C$ field performance compared to the undoped tapes. This can be attributed to the C substitution, which strongly enhances the flux pinning in magnetic fields. Another reason may be related to the improvement of the connection between grains by doping, which increases the current flow between grains. The $J_C$ value of the 10% nano-C doped tapes was lower than that of 5% nano-C doped samples in corresponding magnetic fields. This may be induced by more impurity particles in the 10% doped samples. The existence of relative large amount of impurity particles is believed to suppress intergrain current flow, resulting in a decrease of $J_C$.

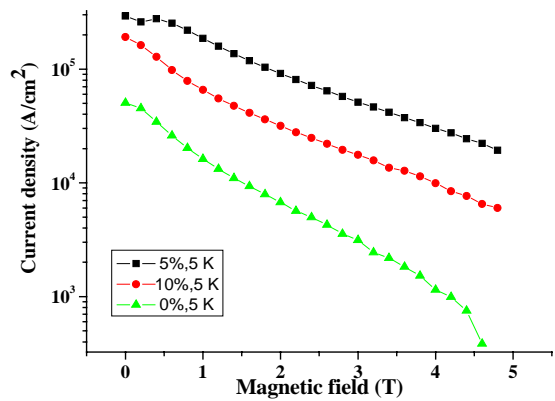

Figure.4. Critical current density as function of the magnetic field at 5 K
for 0%, 5% and 10% nano-C doped samples.

Nano-C doped tapes exhibited a better field performance and much higher values of $J_C$ than undoped samples, suggesting the strongly enhanced flux pinning in magnetic fields, due to the partial substitution of boron for carbon and the partial addition of nano-C particles into a $MgB_2$ matrix. It is also proposed that good grain linkages and grain boundary structures may enhance the $J_C$ values of nano-C doped samples in fields too. However, when the doping level is high (x > 5%), the impurity phases such as nano-C and carbon compound will increases, while the existence of relatively large impurity grains is believed to suppress intergrain current flow, resulting in a decrease of $J_C$. We believe that further improvement in $J_C$-B is expected by either optimizing the doping level or increasing the annealing temperature.

## 4. Conclusions

We have investigated the effect of nano-C addition on the microstructure and superconducting properties of Fe-sheathed $MgB_2$ tapes. Our results demonstrated that both the $J_C$ and flux pinning of $MgB_2$ tapes are significantly enhanced through nano-C doping. The nano-C inclusions and C substitution for B are proposed to be responsible for the enhancement of flux pinning. For the 10 at.% doped samples the $T_C$ dropped only 2.5 K. This role of nanoparticle C may be very beneficial in the fabrication of $MgB_2$ tapes for a large scale of applications.

**Acknowledgments**
The author (Y. Ma) gratefully acknowledges the support of K.C.Wong Education Foundation, Hong Kong. The authors would like to thank Prof. Jiandong Guo, Li Chen, Yan Zhang and Yongzhong Wang in Peking University for their kind help for this work. This work is partially supported by the National Science Foundation of China (NSFC) under Grant No.50472063 and No.50377040 and National "973" Program (Grant No. 2006CB601004).